\documentclass[10pt,conference]{IEEEtran}
\usepackage{cite}
\usepackage{amsmath,amssymb,amsfonts}
\usepackage{algorithmic}
\usepackage{graphicx}
\usepackage{textcomp}
\usepackage{xcolor}
\usepackage[hyphens]{url}
\usepackage{fancyhdr}
\usepackage{hyperref}
\usepackage{algorithm,algorithmic}

\newcommand{\name}{Theseus}
\newcommand{\squishlist}{
 \begin{list}{$\bullet$}
  { \setlength{\itemsep}{0pt}
     \setlength{\parsep}{3pt}
     \setlength{\topsep}{3pt}
     \setlength{\partopsep}{0pt}
     \setlength{\leftmargin}{1.5em}
     \setlength{\labelwidth}{1em}
     \setlength{\labelsep}{0.5em} } }
\newcommand{\squishend}{
  \end{list}  }

\usepackage{titlesec}
\makeatletter
\g@addto@macro{\normalsize}{%
  \setlength{\abovedisplayskip}{3pt plus 0.5pt minus 1pt}
  \setlength{\belowdisplayskip}{3pt plus 0.5pt minus 1pt}
  \setlength{\abovedisplayshortskip}{0pt}
  \setlength{\belowdisplayshortskip}{0pt}
  \setlength{\intextsep}{4pt plus 1pt minus 1pt}
  \setlength{\textfloatsep}{4pt plus 1pt minus 1pt}
  \setlength{\skip\footins}{5pt plus 1pt minus 1pt}
  }
\makeatother

\titlespacing\section{0pt}{2pt plus 1pt minus 1pt}{3pt plus 1pt minus 2pt}
\titlespacing\subsection{0pt}{2pt plus 1pt minus 1pt}{3pt plus 1pt minus 2pt}
\titlespacing\subsubsection{0pt}{2pt plus 1pt minus 1pt}{3pt plus 1pt minus 2pt}

\title{\name: Exploring Efficient Wafer-Scale Chip Design for Large Language Models}

\author{
    \IEEEauthorblockN{Jingchen Zhu\IEEEauthorrefmark{2}, Chenhao Xue\IEEEauthorrefmark{2}, Yiqi Chen\IEEEauthorrefmark{2}, Zhao Wang\IEEEauthorrefmark{2}, Chen Zhang\IEEEauthorrefmark{1}, Yu Shen\IEEEauthorrefmark{2},
    Yifan Chen\IEEEauthorrefmark{2}, Zekang Cheng\IEEEauthorrefmark{4}, \\ Yu Jiang\IEEEauthorrefmark{5}, Tianqi Wang\IEEEauthorrefmark{3}, Yibo Lin\IEEEauthorrefmark{2}, Wei Hu\IEEEauthorrefmark{2}, Bin Cui\IEEEauthorrefmark{2}, Runsheng Wang\IEEEauthorrefmark{2}, Yun Liang\IEEEauthorrefmark{2}, Guangyu Sun\IEEEauthorrefmark{2}}
    \IEEEauthorblockA{\IEEEauthorrefmark{2} Peking University  \IEEEauthorrefmark{1} Shanghai Jiao Tong University \IEEEauthorrefmark{3} Huawei}
    \IEEEauthorblockA{\IEEEauthorrefmark{4} University of Science and Technology of China \IEEEauthorrefmark{5} Tsinghua University}
    
    \IEEEauthorblockA{\{zjc990112, gsun\} @pku.edu.cn}
}

\begin{document}
\maketitle

\ifdefined\hpcacameraready 
  \thispagestyle{camerareadyfirstpage}
  \pagestyle{empty}
\else
  \thispagestyle{plain}
  \pagestyle{plain}
\fi

\newcommand{\hpcaheight}{0mm}
\ifdefined\eaopen
\renewcommand{\hpcaheight}{12mm}
\fi

\begin{abstract}
The emergence of the large language model~(LLM) poses an exponential growth of demand for computation throughput, memory capacity, and communication bandwidth. Such a demand growth has significantly surpassed the improvement of corresponding chip designs. With the advancement of fabrication and integration technologies, designers have been developing Wafer-Scale Chips~(WSCs) to scale up and exploit the limits of computation density, memory capacity, and communication bandwidth at the level of a single chip. 
Existing solutions have demonstrated the significant advantages of WSCs over traditional designs, showing potential to effectively support LLM workloads.

Despite the benefits, exploring the early-stage design space of WSCs for LLMs is a crucial yet challenging task due to the enormous and complicated design space, time-consuming evaluation methods, and inefficient exploration strategies. To address these challenges, we propose \name, an efficient WSC design space exploration framework for LLMs. We construct the design space of WSCs with various constraints considering the unique characteristics of WSCs. We propose efficient evaluation methodologies for large-scale NoC-based WSCs and introduce multi-fidelity Bayesian optimization to efficiently explore the design space. Evaluation results demonstrate the efficiency of \name\ that the searched Pareto optimal results outperform GPU cluster and existing WSC designs by up to 62.8\%/73.7\% in performance and 38.6\%/42.4\% in power consumption for LLM training, while improving up to 23.2$\times$ and 15.7$\times$ for the performance and power of inference tasks. Furthermore, we conduct case studies to address the design tradeoffs in WSCs and provide insights to facilitate WSC designs for LLMs. 

\end{abstract}

\section{Introduction}


Recent advancements in neural networks have led to a significant escalation in model size, particularly in the realm of large language models (LLMs). The evolution from Bert to GPT4~\cite{bert-1, llama, palm, gpt3-1, openai2023gpt4} has resulted in a parameter increase exceeding a thousand-fold. This upward trend is anticipated to persist, owing to the superior performance of larger models in tasks related to natural language understanding and content generation~\cite{touvron2023llama,t5,opt}. Concurrently, it necessitates significantly enhanced computational throughput, memory capacity, and data communication bandwidth in hardware infrastructures~\cite{rajbhandari2021zero, gshard, taccl, dt-bw}. The surge in demand has markedly exceeded advancements in corresponding chip designs like GPUs~\cite{A100, H100}, attributed to the field size limitation of lithography steppers. This constraint is referred to as the {\bf reticle limit}, defined as 26 mm by 33 mm, or 858 mm²~\cite{reticle_limit}. 
State-of-the-art chip designs, such as NVIDIA H100~\cite{H100}, B200~\cite{B200}, and Intel Gaudi3~\cite{gaudi3}, are approaching and even surpassing the reticle limit in scale.

Fortunately, advancements in chiplet design method~\cite{shao2019simba}, die-to-die interconnection~\cite{turner2018ground}, and 2.5D/3D stacking methods~\cite{loh2007processor}, have facilitated scaling beyond the reticle limit. Innovative chip designs that surpass the limit have been proposed for artificial intelligence applications~\cite{ponte-vecchio, mi200, lie2022cerebras, chang2022dojo}, etc. Among these solutions, the {\bf Wafer-Scale Chip (WSC)} designs (e.g. Cerebras WSE2 and Tesla's Dojo) stand out as a promising approach to maximize computation density, on-chip memory capacity, and communication bandwidth. Cerebras WSE2~\cite{lie2022cerebras} integrates 850,000 cores, 40GB of on-chip SRAM, and 200Pb/s of on-chip fabric bandwidth on a 46,225mm² monolithic silicon substrate and achieves 7.5 PFLOPS for large-scale GEMM operation. Tesla's Dojo~\cite{chang2022dojo} comprises 25 D1 dies, 10TB/s on-tile bisection bandwidth, and 36TB/s off-tile aggregate bandwidth, achieving a performance of 9 PFLOPS. It is worth mentioning that WSCs generally refer to efforts to scale up a single chip with on-wafer interconnections, rather than solely large-scale chip designs that reach the wafer area limit. For instance, Tesla's Dojo chip occupies only one-third of a 12-inch wafer. Our exploration results also indicate that reaching the physical limit is often not the optimal design choice for WSCs when considering LLM workloads. Compared to GPUs, these WSC designs offer over $7\times$ the peak performance, $200\times$ on-chip memory bandwidth, and $5\times$ the inter-chip bandwidth, alongside significantly improved energy efficiency, enabling more effective scaling for both training and inference tasks on LLMs.


Despite the aforementioned advantages of WSCs, it is crucial to determine the optimal configurations of WSCs to strike a balance in utilizing diverse resources, achieve peak performance, and enhance energy efficiency. Our experiments have demonstrated that improper designs can significantly degrade the achievable performance of WSCs in LLM workloads, sometimes by several tens of times. Furthermore, varying application workloads and optimization objectives can result in a range of designs, each with notable differences. Hence, early-stage \textbf{Design Space Exploration~(DSE)} becomes imperative in crafting efficient WSCs capable of delivering optimal performance and energy efficiency across a spectrum of application requirements. 

A Design Space Exploration (DSE) process typically involves three key stages: \textbf{design space construction}, \textbf{design point evaluation}, and \textbf{exploration strategy}. Although DSE has been thoroughly studied on AI accelerators~\cite{chen2019eyeriss, gao2019tangram, sparseloop, maestro, timeloop, interstellar}, adopting existing DSE methods to the WSC scenario introduces challenges in all these stages. Firstly, the design space of WSCs is larger and more complex, requiring consideration of numerous parameters and various constraints compared to traditional AI accelerators. Secondly, there exists several obstacles to evaluating the performance of a WSC configuration accurately and efficiently, especially with regard to the communication behaviours at scale. Lastly, conducting multi-objective optimization over the complex design space poses challenges for the explorer to efficiently characterize the target space and find optimal designs with a minimum number of sampled design points. To tackle these challenges, we propose \name, a DSE framework to facilitate high-efficiency WSC design for LLM workloads. The contributions of this paper can be summarized as follows:

\squishlist
    \item We construct an extensive design space that encompasses WSC architecture configurations with wide range of candidate values, and introduce parameters to explore heterogeneous WSC designs. We consider various design constraints including area, power, yield, stress, etc. Specifically, to address yield constraints, we extend the existing yield model and introduce redundancy-based yield enhancement to the WSC scenario.
    \item We propose a hierarchical methodology for evaluating LLMs on WSCs through tile-level, op-level and chunk-level evaluations with considerations of communication at different levels. At the op-level, we employ a graph neural network~(GNN)-based method for fast and accurate NoC estimation to support multi-fidelity optimization. 
    \item We propose a multi-fidelity multi-objective Bayesian optimization~(MFMOBO) algorithm to efficiently explore the design space of WSCs for LLMs and leverage the advantages of various evaluation methodologies.
    \item We conduct case studies using our proposed \name\ framework for addressing the design tradeoffs in WSCs, and provide analysis and insights to facilitate the design optimization of WSCs for LLMs.

\squishend

\section{Background}

\subsection{WSC Basics}

Wafer-Scale Chip~(WSC) design has emerged as a promising solution to alleviate inter-reticle communication overhead with low-cost on-wafer interconnections and has demonstrated advantages in computational power and on-chip memory capacity for efficiently supporting neural network workloads. The concept of WSCs has attracted significant attention from both industry and academia. Recently, two notable commercial WSC designs have been proposed. Cerebras WSE2~\cite{lie2022cerebras} employs offset exposures and proprietary layers of interconnect to stitch dies together directly on monolithic wafers. This results in a uniform and continuous fabric across the reticle boundary, which provides good hardware abstraction. Tesla Dojo~\cite{chang2022dojo} uses a redistribution layer (RDL) to interconnect D1 chips with SerDes and integrate D1 chips with integrated fan-out system-on-wafer~(InFo-SoW) packaging, with known-good-die (KGD) techniques to ensure yield requirement. Previous research has discussed the methodology and challenges of WSC design~\cite{pal2019architecting, huang2016efficient, pal2021designing}, as well as efficient workload mapping on specific WSC engines~\cite{liu2022partition, lin2022mapping, li2021placement}.

\subsection{LLM training and inference}

LLM training tasks are characterized by numerous GEMM operations, which heavily rely on high computational power. In distributed training systems, the communication bandwidth also significantly influences training performance, as there is frequent data exchange and parameter updates between chips and nodes. Additionally, LLM training tasks require high memory capacity to store parameters and intermediate results. 

LLM inference tasks consist of two stages: the prefill stage and the decode stage. The prefill stage exhibits similar characteristics to the forward pass of training, involving compute-intensive operations. In contrast, the decode stage mainly consists of small-batch GEMV operations and Key-Value~(KV) cache memory access due to the autoregressive decoding process in LLMs. Therefore, the decode stage places high demands on memory bandwidth. The diverse requirements of these two stages pose a challenge for accommodating them with the same hardware design, motivating exploration into heterogeneity in WSCs to effectively support LLM inference.

Prior studies~\cite{shoeybi2019megatron, zheng2022alpa, li2023alpaserve} have delved into the exploration of parallel strategies by employing model partitioning along various dimensions. Three practical and widely utilized parallel strategies for scaling deep learning systems for LLM workloads are proposed, including \textit{data parallel (DP)}, \textit{pipeline parallel (PP)}, and \textit{tensor parallel (TP)}. For LLMs featuring evenly-sized stages, exploration results suggest that the application of these three commonly employed parallelism techniques can yield commendable scalability~\cite{zheng2022alpa}. 



\subsection{WSC for LLM: Advantages and Challenges}

The large computational power and on-chip buffer of WSCs provide evident advantages for LLM workloads, facilitating enhanced performance for compute-intensive operations and reduced off-chip memory access through efficient on-chip data reuse. Efficient inter-reticle communication is also pivotal for overall performance and power improvements. In both LLM training and inference tasks, the ever-growing model sizes necessitate the adoption of efficient parallel strategies~\cite{shoeybi2019megatron, aminabadi2022deepspeed} and memory optimizations~\cite{rajbhandari2020zero, rajbhandari2021zero} to meet memory capacity constraints in distributed systems. 
These optimizations typically increase the communication between chips and nodes, which can benefit greatly from the inter-reticle connections of WSCs. Additionally, WSCs can leverage stacking memory with high memory bandwidth to meet the demands of the decode stage during inference. Although high bandwidth may necessitate sacrificing some memory capacity, the introduced communication can still be efficiently managed by inter-reticle communication to ensure improved overall performance. However, compared to scale-out solutions, WSCs have less flexible interconnect topologies between reticles, often constrained to 2D-mesh architectures. This poses challenges for both communication performance and system evaluation at scale.

\section{Motivational Analysis}

\noindent\textbf{Challenge 1: Design Space Construction. } In addition to the basic architecture parameters in traditional accelerator chips, WSCs introduce design considerations at both the reticle and wafer levels. Due to the large scale of WSCs, design configurations such as computational capacity, memory organization and interconnection bandwidth exhibit a broader range of variability. Meanwhile, WSCs need to validate against various design constraints. To address this challenge, in \name, we construct an extensive design space of WSCs for LLMs, detailed in Sec.~\ref{space}, and propose modeling approaches for various WSC metrics including area, power, yield and stress.

\noindent\textbf{Challenge 2: Design Point Evaluation. }
With the expansive scale of computational resources in WSCs, the data transfer between cores and reticles emerges as a crucial concern of system performance. Due to the complexity of transmissions introduced by multi-level parallelisms when mapping LLMs onto WSCs, as well as the scheduling and mapping of task Directed Acyclic Graphs (DAGs) onto local spatial architectures, we must carefully consider traffic flow congestions in NoC estimation. Traditional DSE methods rely on cycle-accurate~(CA) simulators~\cite{booksim} for NoC evaluation, which can be time-consuming at the scale of WSCs. Existing NoC analytical models, in contrast, can provide rapid NoC estimation results. However, these models usually assume predetermined traffic patterns~\cite{DBLP:journals/todaes/QianBTM16, DBLP:journals/tcad/OgrasBM10} (e.g. tornado or uniform), which may not capture the application-specific details. Recently, several machine learning-based methods have been proposed to efficiently evaluate application-specific NoC performance~\cite{qian2013svr, li2022noception}. However, these methods face limitations in dealing with variable package sizes and the heterogeneity in NoC bandwidth, thereby impeding their direct application to WSC scenarios. To address this challenge, in \name, we propose a hierarchical evaluation method to reduce the estimation scale of NoC, and introduce a GNN-based method for fast and accurate NoC performance estimation. 

\noindent\textbf{Challenge 3: Exploration Strategy. } WSC designs are designed with stringent power constraints determined by the design of heat dissipation and power delivery network
~\cite{pal2019architecting}. This necessitates a delicate balance between optimizing both computation performance and power consumption. The multi-objective optimization considering both performance and power of WSCs is non-trivial, especially within our design space with enormous design parameters, and irregular shapes due to various design constraints. Meanwhile, enhancing the convergence and efficiency of this optimization process to minimize the number of iterations requires careful algorithm design. To address this challenge, in \name, we introduce MFMOBO to leverage the information from less accurate yet faster evaluation methodologies.

\section{\name\ Overview}

\begin{figure}[t]
  \centering
  \includegraphics[width=0.9\columnwidth]{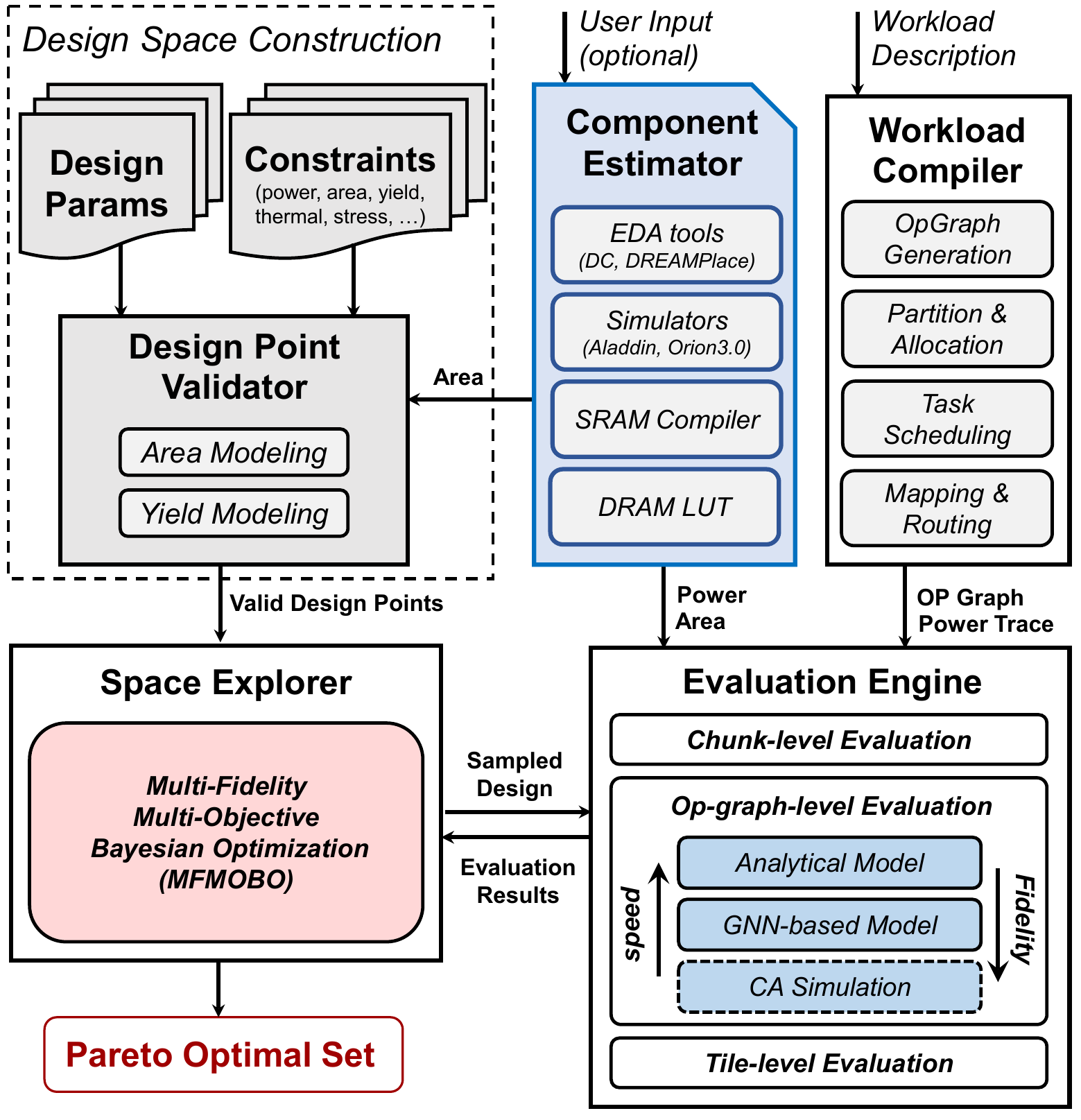}
  \caption{\name\ Framework Overview}
  \label{fig:dse_framework}
\end{figure}

To tackle the challenges mentioned above, we propose \name, an efficient design space exploration framework that jointly optimizes performance and power for WSC to search for Pareto-optimal designs.
Fig.~\ref{fig:dse_framework} presents an overview of \name. The DSE process begins with design space construction, which involves combining the design parameters of WSC and employing the \textbf{Design Point Validator} to discard design points that can not satisfy the constraints. Next, the \textbf{Space Explorer} employs our proposed Multi-Fidelity Multi-Objective Bayesian Optimization~(MFMOBO) to iteratively sample design points from the constructed design space. To evaluate the sampled design points, we employ a \textbf{Component Estimator} to calculate the area and power of WSC basic modules, and implement a \textbf{Workload Compiler} to compile LLM workloads onto WSCs. The \textbf{Evaluation Engine} 
hierarchically evaluates the selected WSC design in tile-level, op-graph-level and chunk-level evaluation. 
The evaluation engine provides performance and power results in different fidelity according to the needs of the explorer. Finally, the explorer selects design points for the next iteration based on the feedback from the evaluation engine.  This iterative optimization process continues until reaching the pre-set iterations number, and the \name\ framework outputs the searched Pareto optimal set. In the following sections, we detail the designs of individual components within \name.


\section{WSC Design Space}\label{space}

\subsection{WSC Architecture Parameters}\label{subsec:arch_para}

\begin{figure}[t]
  \centering
  \includegraphics[width=\columnwidth]{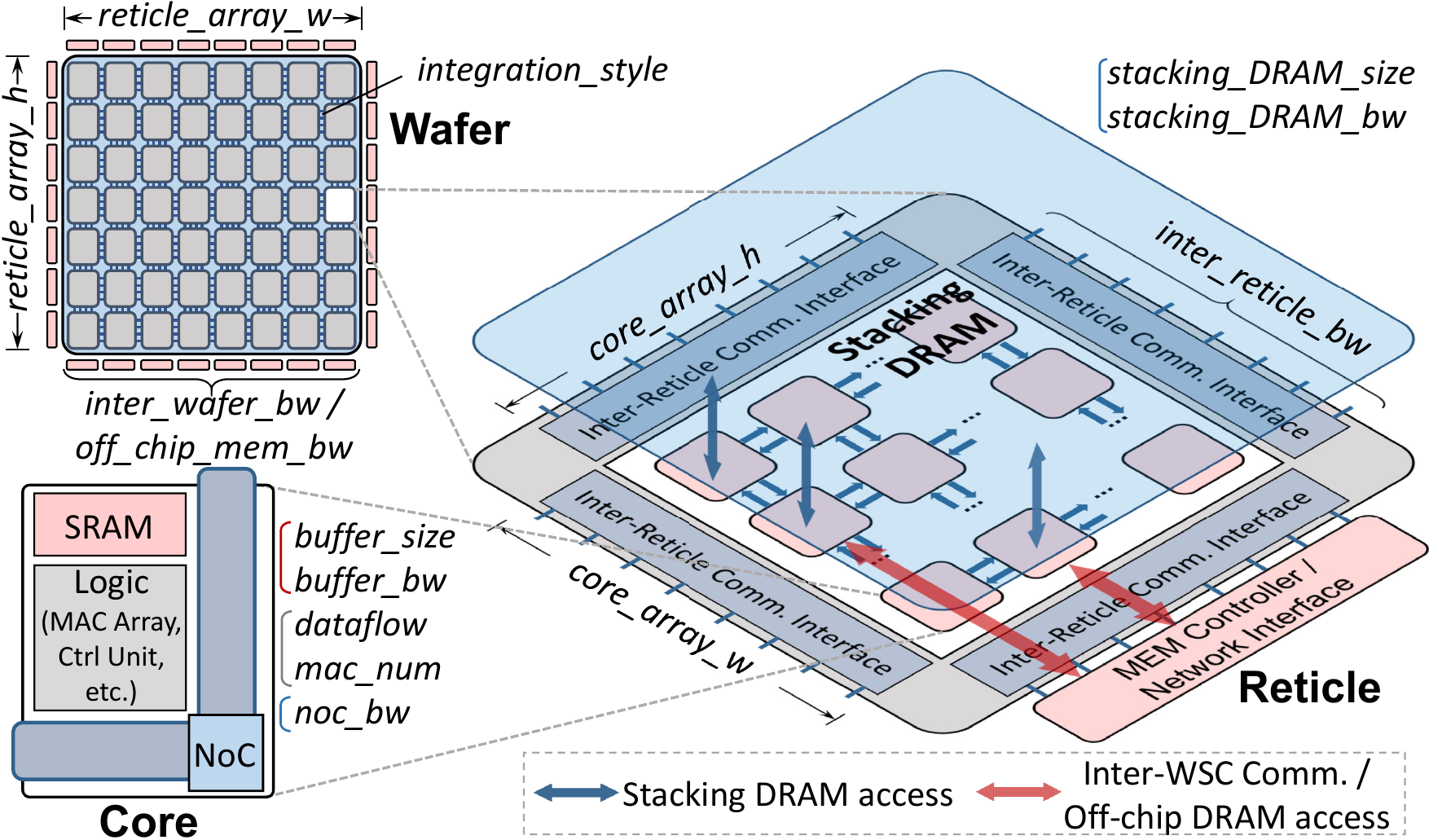}
  \caption{Typical WSC Design and Architecture Parameters}
  \label{fig:wsc-arch}
\end{figure}

To comprehensively explore the architecture design of WSCs, we consider crucial parameters in the three hierarchies of core, reticle and wafer. Fig.~\ref{fig:wsc-arch} labels these architecture parameters. At the \textbf{Core} level, on-chip SRAM capacity impacts the core's data reuse ability and the granularity of communication between cores. SRAM bandwidth determines the utilization of processing units and impacts intra-core dataflow optimization. Dataflow describes the pattern of data transmission and reuse across MAC units and the memory hierarchy, shaping the design of fixed datapaths and the operational efficiency of various operators~\cite{shao2019simba, gao2019tangram, du2015shidiannao}. MAC number sets the upper limit of a single core's tensor computation capacity. Lastly, NoC bandwidth is pivotal in determining the effectiveness of inter-core communication. 
At the \textbf{Reticle} level, cores are connected with NoC to form a 2D-mesh array, which impacts the shape and peak performance of a reticle. Inter-reticle communication bandwidth are provided by the communication interfaces around the reticle to support data transmission across the reticle boundaries. We also consider stacking DRAM for efficient memory access based on through-silicon-via~(TSV), with a certain capacity and bandwidth per unit area. 
At the \textbf{Wafer} level, reticles are further integrated into an array of certain height and width. The integration technology determines the area overhead and power consumption for inter-reticle communication. The memory controllers and network interfaces around the wafer provide off-chip DRAM access bandwidth and communication bandwidth between WSCs, which allows larger memory capacity and further scaling out of WSC systems.



\subsection{Heterogeneous Modeling}

\begin{figure}[t]
  \centering
  \includegraphics[width=\columnwidth]{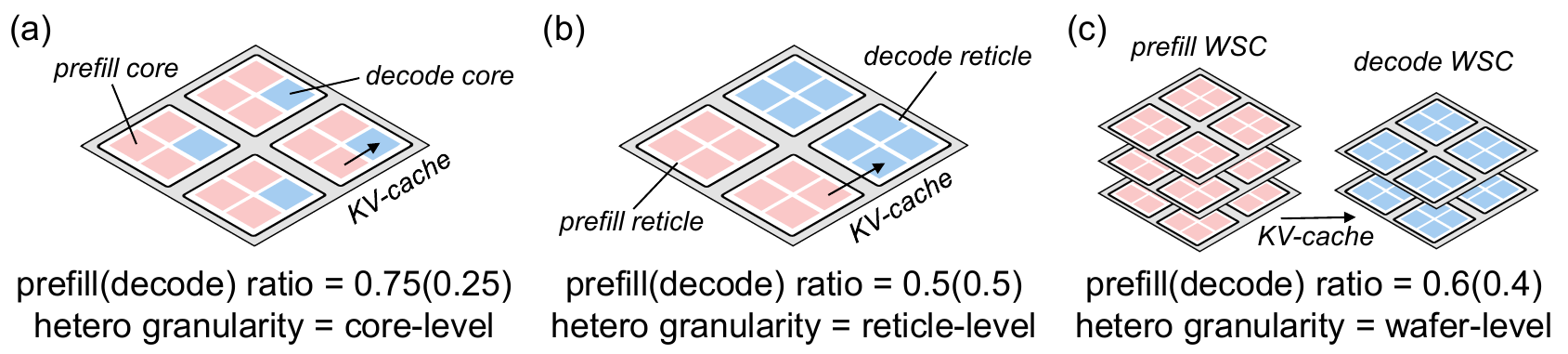}
  \caption{Examples of Heterogeneous Modeling}
  \label{fig:heterogeneous}
\end{figure}

To investigate the impact of heterogeneous WSC design on LLM inference performance, we introduce two parameters for characterization: prefill(decode) ratio and heterogeneous granularity. The prefill(decode) ratio represents the proportion of computational resources allocated to the prefill and decode stages. Heterogeneous granularity indicates the level of heterogeneity in the architecture hierarchy. Fig.~\ref{fig:heterogeneous} presents several examples of heterogeneous design at different levels. Core-level heterogeneity is achieved through software scheduling, where different stacking memory bandwidths are allocated to prefill and decode stage cores within the same reticle. Reticle-level and wafer-level heterogeneities are achieved by adjusting stacking memory bandwidth. In reticle-level heterogeneity, heterogeneous reticles with different memory bandwidths are integrated into the same WSC and simultaneously execute both stages, while wafer-level heterogeneity requires computations for the prefill stage and decode stage to be performed on different WSCs.

\subsection{Defective Core Modeling}



Traditionally, a core's yield is influenced by its area and the process technology employed. The relationship is encapsulated by the well-known Murphy Model~\cite{murphy1964cost}, as shown in Equation~\ref{eq:murphy}. In this model, the parameter $A$ represents the core area in $cm^2$. Meanwhile, $D_{0}$ denotes the average defect density, measured in defects per $cm^2$, and serves as an empirical parameter associated with the process technology. 
\begin{equation}
    Yield_{Murphy}= \left[ \frac{1-e^{-AD_{0}}}{AD_{0}} \right]^{2}
    \label{eq:murphy}
\end{equation}

\begin{figure}[t]
  \centering
  \includegraphics[width=0.8
  \columnwidth]{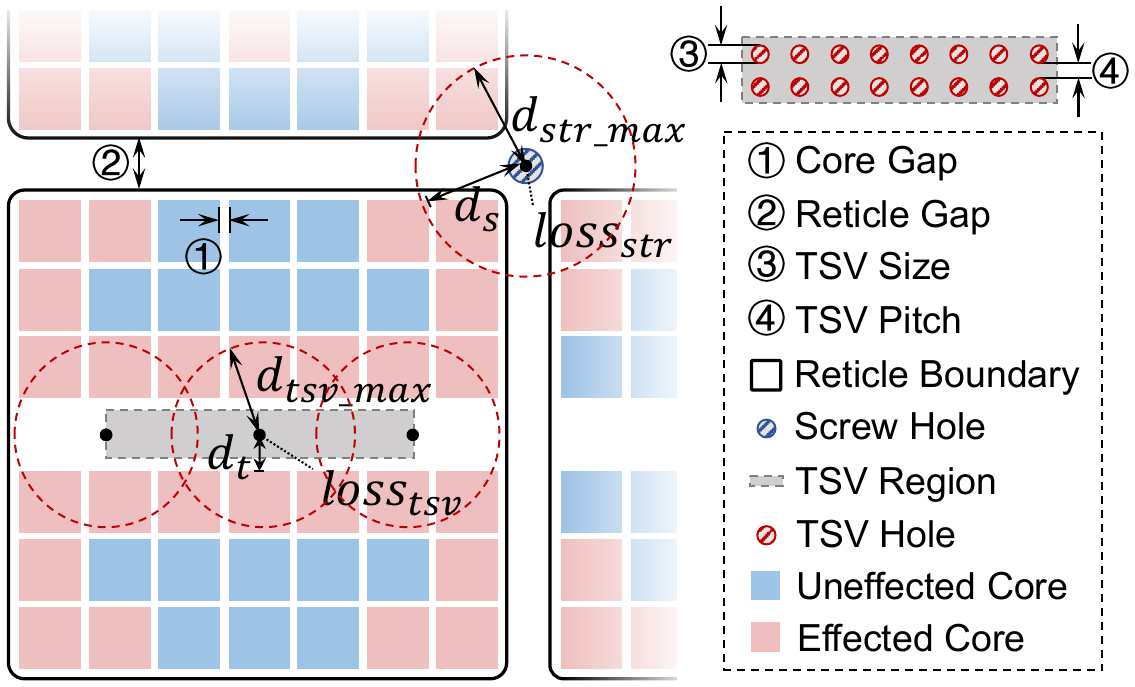}
  \caption{Impact of Screw Holes and TSVs on Core Yield}
  \label{fig:defect}
\end{figure}

In WSC design, additional factors such as screw holes and Through-Silicon Vias (TSVs) may further impact a core's yield. To maintain the wafer's flatness and stability during manufacturing, it is often secured to the underlying PCB using screws. Screw holes are strategically placed at the intersections of reticles. The stress exerted by these screws can lead to a certain degree of yield degradation in the areas surrounding the holes. Similarly, TSVs, which involve drilling holes in the center of the reticle and filling them with conductive material for electrical interconnections between silicon layers, can also result in yield loss for nearby cores. The impact of screw holes and TSVs on yield is depicted in Fig.~\ref{fig:defect}.

As depicted in Fig.~\ref{fig:defect}, the symbol $d_{str\_max}$ represents the maximum distance over which a stress hole influences yield. The effect of a stress hole on yield decreases linearly with distance, as estimated in Equation~\ref{eq:stress}. In this equation, $d_{s}$ signifies the distance from the stress hole to the nearest vertex of the core, while $loss_{str}$ indicates the rate of yield loss attributed to the center of the stress hole.
\begin{equation}\label{eq:stress}
     Yield_{str}=\left[\frac{loss_{str}}{d_{str\_max}}d_{s}+1-loss_{str}\right]
\end{equation}

The model describing the impact of TSVs on yield closely mirrors that of screw holes. TSVs must be created on the wafer's surface due to their functional requirements and inherent characteristics. Consequently, the area designated for TSVs cannot overlap with the area allocated for a core. The additional overhead is linked to the total number of TSVs, which in turn is dictated by the bandwidth requirements of stacked DRAM. The yield of cores situated within a distance less than $d_{TSV\_max}$ from a TSV is affected, and this can be quantified similarly to Equation~\ref{eq:stress}.


In conclusion, the yield of a core can be determined by consolidating all these elements, as outlined in Equation~\ref{eq:yield}. 
\begin{equation}\label{eq:yield}
    Yield_{core} = Yield_{Murphy} \times Yield_{str} \times Yield_{TSV}
\end{equation}

\subsection{Core Redundancy} 
With the error model of an individual core established, we can extend the calculation to determine the yield at both the reticle and wafer levels. Considering the substantial drop in overall yield that may occur when integrating numerous cores onto a single wafer, implementing appropriate redundancy mechanisms becomes essential to attain an acceptable yield target. 
A commonly employed strategy is to designate a portion of the cores as redundant, enabling them to substitute defective cores as required~\cite{lie2022cerebras}. Assuming a reticle contains $p$ operational cores and $n$ redundant cores, the yield at the reticle level can be computed using Equation~\ref{redundancy_cal}, where $Y_{core}$ denotes the yield of individual cores within the reticle. Similarly, the wafer level yield can be further calculated using the reticle level yield. 
\begin{equation}
Y_{PS} = \sum_{i=p}^{p+n}{\binom{p+n}{i}{Y_{core}}^{i}{(1-Y_{core})^{p+n-i}}}
\label{redundancy_cal}
\end{equation}

In setting a yield target, the selection of the percentage of redundant cores is critical, as this can lead to increased area overhead and a reduction in effective computational capability. This decision can also impact the selection of architectural parameters during DSE.
Furthermore, it is crucial to acknowledge that the choice of integration technique during DSE significantly influences yield. For example, Dojo's adoption of the known-good-die technique may lead to higher yields compared to Cerebras's die-stitching method. The disparity in these approaches can influence the overall yield outcomes, as well as the DSE results.

\subsection{Design Constraints}

Prior to initiating the exploration process, it is prudent to establish appropriate design constraints. This step helps in precluding invalid design solutions and enhancing the overall efficiency of the exploration. Key constraints that should be considered are discussed in this subsection.

%
\noindent{\textbf{Area Constraint:}} The area of the reticle and wafer cannot exceed the limits of the lithograph and silicon substrate.


\noindent{\textbf{Power Constraint:}} The operating power of WSC cannot exceed the pre-determined power limit. 

\noindent{\textbf{Yield Constraint:}} Considering the reserved redundant cores, WSC needs to meet the yield requirement of the design. 

\noindent{\textbf{SRAM Constraint:}} Some combinations of SRAM configurations are infeasible from the SRAM Compiler.

\noindent\textbf{Stress Constraint:} The area for stacking DRAM TSV holes should be less than the pre-determined area ratio~(e.g., 1.5\%) of a reticle for stress considerations.

\section{Evaluation Methodology}


\begin{figure}[t]
  \centering
  \includegraphics[width=\columnwidth]{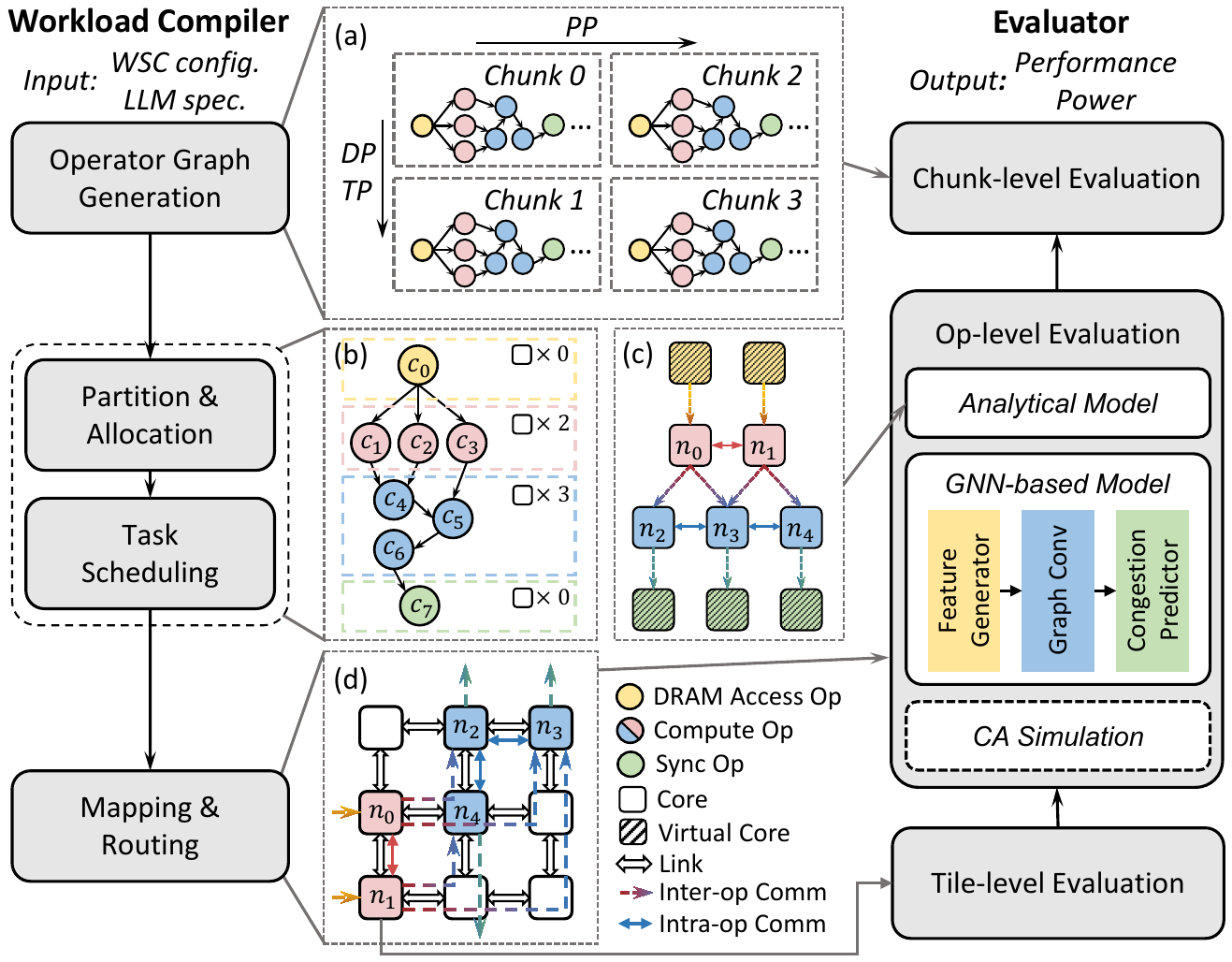}
  \caption{Overview of the Evaluation Methodology}
  \label{fig:evaluation}
\end{figure}

Fig.~\ref{fig:evaluation} illustrates the overview of evaluation methodology in \name. Specifically, we employ a Workload Compiler to determine the operating pattern and generate essential evaluation information. Based on this, we propose a hierarchical methodology to reduce the scale of evaluation, which estimates the performance and power through tile-level, op-level and chunk-level evaluation. 

\subsection{Workload Compiler}~\label{sec:compile}

As shown in Fig.~\ref{fig:evaluation}, Workload Compiler takes the WSC configuration and benchmark workload as inputs, and runs in the following steps: (1) \textbf{Operator Graph Generation:} Given a parallel strategy, the LLM model is first segmented into model chunks. Workload Compiler generates the operator graph for each chunk, as well as the data transmission between model chunks. Meanwhile, Workload Compiler divides all compute resources evenly according to the number of chunks, and binds each chunk to the corresponding compute resource. (2) \textbf{Partition and Allocation:} Workload Compiler partitions the operator graph of each chunk into disjoint subgraphs, allocating resources to enable efficient intra-op and inter-op parallelism according to the methodologies in prior works~\cite{timeloop, gao2019tangram, cai2023inter}, while considering the SRAM capacity limits. (3) \textbf{Task Scheduling:} Workload Compiler divides the operators in subgraphs into tiles according to the allocated number of logic cores and generates communication traces at the level of cores. (4) \textbf{Mapping and Routing:} Workload Compiler maps logic cores to the physical core array and generates the routing paths for data transmission between cores. During the workload compiling process, we iterate through all combinations of TP, DP, PP, and micro-batch sizes that satisfy the memory capacity constraint and select the best-performance parallel strategy based on the evaluation results.




\subsection{Tile-level Evaluation}

Tile-level evaluation focuses on the evaluation of tensor operations on cores with fixed dataflow, which have been extensively studied and analyzed in prior works~\cite{timeloop, maestro}. Similarly, we perform unrolling and tiling over specific loop dimensions for tile-level evaluation, while considering the impact of SRAM capacity on data reuse. We record the interval of the tiled output for subsequent NoC estimation.

\subsection{Op-level Evaluation}
Based on the latency and communication patterns of individual cores from tile-level evaluation, op-level evaluation further estimates the performance of NoC-based core array with both inter-operator and intra-operator communications~\cite{cai2023inter}. In op-level evaluation, we propose two methodologies trading off evaluation speed and accuracy for NoC estimation under complex communication scenarios, including the analytical model and GNN-based model.

\noindent\textbf{Analytical Model.} 
For fast estimation at the early iterations of DSE, we propose an analytical model for the communications on the core array within a chunk. Based on the output of Workload Compiler, we figure out the data transmission volume on each link of the NoC and calculate the communication time between cores with the equivalent bandwidth. We label the communication delays on each edge of the logic core graph~(Fig.~\ref{fig:evaluation}(c)) and traverse the graph in topological order to select the longest path as the overall latency of the chunk. It is worth mentioning that the actual latency of a chunk needs to take into account the overhead of DRAM access and inter-chunk synchronization, which correspond to the edges connecting virtual cores in Fig.~\ref{fig:evaluation}(c). These overheads will be considered in the chunk-level evaluation.

\noindent\textbf{GNN-based Evaluation.} To enhance the accuracy of NoC estimation by considering the congestion, we propose a novel GNN-based performance evaluation method. Compared to existing machine learning methodologies, our model effectively addresses the issue of variable packet sizes and the inherent heterogeneity in NoC bandwidth to support the evaluation of LLM workloads on WSCs.


Our GNN model takes the core topology graph (Fig.~\ref{fig:evaluation}(d)) generated from Workload Compiler as input, which provides the inter-link topological relationships for GNN to characterize NoC congestion. Meanwhile, we encode the packet injection rate and transmission volume in the feature matrix of nodes and edges in the topology graph, respectively. 

Fig.~\ref{fig:evaluation} presents the architecture of our GNN model. \textit{Feature generator} consists of MLPs that project node features $x_v$ and edge features $x_e$ to initial hidden states $h_v^0$ and $h_e^0$. \textit{Graph Convolution module} aggregates neighboring information with message passing mechanism \cite{gilmer2017neural} and update nodes' hidden states for $T$ iterations. Similar to \cite{li2022noception}, message passing is conducted on both the original graph $G$ and its reversed graph $\tilde{G}$ to model both upstream contention and downstream backpressure. \textit{Congestion predictor} predicts average channel waiting time $y_e$, which can reflect the traffic pattern of NoC. For edge $e = (u, v)$ that represents a physical link, $y_e$ is predicted by
\begin{equation}
    \hat{y}_e = \theta(\text{Concat}(h_u^T, h_v^T, h_e^0)),
\end{equation}
in which $\theta$ is a MLP. With the predicted average channel waiting time $\hat{y}$, we can reconstruct end-to-end transmission latency between cores. For a packet with $k$ flits, the average transmission latency $t(k)$ can be calculated by:
\begin{equation}
    t(k) = k + \sum_{\{l | \varphi(l) = 1\}} \hat{y}_l, 
\end{equation}
in which $\varphi(l)$ is a Boolean indicating whether the transmission utilizes link $l$. We apply the GNN model on all edges between cores in the logic core graph~(Fig.~\ref{fig:evaluation}(c)). Similar to the analytical model, we can reconstruct the overall latency of a chunk by finding the critical path of the graph.

\noindent\textbf{Cycle-accurate Simulation.} We extend BookSim2~\cite{jiang2013detailed} for cycle-level simulation and dataset generation. We design a series of instructions and micro-instructions to describe the compute, memory access and communication of WSC cores, and connect the cores to Booksim routers. We add instruction support for Booksim to inject and forward packets according to the micro-instructions sent from the network interface of cores. We argue that for accelerator cores, when dealing with regular tensor operations, the latency for computation and memory access is relatively deterministic. So we simplify the estimation of computation and memory access latency inside cores by analytical models. Specifically, 2D-DRAM accesses are considered to include local memory accesses of controllers, and data communication from memory controllers to target cores. During the compilation and generation of instructions, we consider the data dependency of computational operations and the memory capacity constraints.

\subsection{Chunk-level Evaluation}

Chunk-level evaluation further considers the data transfer between chunks, including TP-induced collective communication, PP-induced cross-pipeline-stage communication, and DP-induced weight update communication. Additionally, we consider the traffic related to traditional off-chip DRAM access at the chunk level. Despite the equitable partitioning of chunks in terms of workload and hardware resources, the overhead for off-chip DRAM access can be different due to the diverse locations of reticles assigned to individual chunks. Considering all inter-chunk communications and DRAM access demands, we count the amount of data transmission for each inter-reticle link and assess the communication latency based on the available bandwidth. The DRAM access latencies are then combined with op-level evaluation results to derive the overall performance of individual chunks. Meanwhile, we consider the pipeline efficiency based on micro batch size to calculate the throughput on target workloads.

\subsection{Area and Power Estimation}


\noindent\textbf{Area Estimation.} We utilize the SRAM compiler to generate various configurations of SRAM Macros along with their height and width. We also implement MAC array in different dataflows, NoC routers, and RISC-V core as the control unit with Chisel based on Purlin~\cite{guo-purlin-iccd2022}. The generated RTL is then synthesized using Synopsys Design Compiler, and the netlist is further input to DREAMPlace~\cite{lin2020dreamplace} for placement to obtain the area of the core array within a reticle. In estimating the inter-reticle distance and area overhead of inter-reticle communications, we refer to Cerebras WSE2, Tesla Dojo, and Nvidia GRS~\cite{turner2018ground} for relevant information.

\noindent\textbf{Power Estimation.} 
We adopt a methodology similar to Aladdin~\cite{shao2014aladdin}, where we define a series of actions, including MAC operations, NoC transmissions, inter-reticle communications, and SRAM/DRAM accesses, etc., to estimate power consumption. We utilize SRAM Compiler to obtain SRAM static power and energy consumption for read and write operations. We refer to Aladdin, Orion3.0~\cite{kahng2015orion3}, as well as existing chiplet and WSC designs~\cite{nvdla, tan2021nn, lie2022cerebras, pal2019architecting, turner2018ground} for the power estimation of computational logic, NoC components, and communication modules.  During the evaluation process, we record the number of action calls to calculate dynamic power and combine it with the static power of components to estimate power consumption for WSC designs. 



To reduce the time required for evaluating each sampled design, we develop the Component Estimator to construct a dataset containing the area and power of WSC basic modules with various configurations. During the DSE process, we use evaluation models to combine the results obtained from the Component Estimator and calculate the area and power of WSCs. This area-power table can be updated with more precise results as required, facilitating efficient and precise evaluations of the design.

\section{Space Explorer Design}

To efficiently explore the design space of WSCs for LLMs, we propose a multi-fidelity multi-objective Bayesian optimization~(MFMOBO) algorithm. By leveraging the informative nature of low-fidelity objective functions to enhance high-fidelity optimization, MFMOBO can attain faster convergence and better results, especially in the early iterations.


\begin{algorithm}[]
    \renewcommand{\algorithmicrequire}{\textbf{Input:}}
	\renewcommand{\algorithmicensure}{\textbf{Output:}}
	\caption{Pseudo-code for the MFMOBO Algorithm}
    \label{alg:mfmobo}
    \begin{algorithmic}[1] 
        \REQUIRE $\mathbb{A}$, $f_0$, $f_1$, $d_0$, $d_1$, $k$, $N_0$ and $N_1$; 
        \STATE Init the prior of $f_0$: $\mathbb{D}_0\leftarrow sample(f_0, \mathbb{A}, d_0)$ 
        \STATE Init the prior of $f_1$: $\mathbb{D}_1\leftarrow sample(f_1, \mathbb{A}, d_1)$ 
        \STATE $(\mathbb{D}, \mathbb{M}, f, \chi)\leftarrow (\mathbb{D}_1, \mathbb{M}_1, f_1, \chi_1)$
        \STATE \textbf{for} $i \leftarrow 0$ until  $N_0+N_1-d_0-d_1$ \textbf{do}
        
        \STATE \quad \textbf{if} $i = N_1-d_1$ \textbf{do}
        \STATE \quad \quad $(\mathbb{D}, f, \chi)\leftarrow (\mathbb{D}_0, f_0, \chi_0)$
        
        \STATE \quad \textbf{if} $i = N_1-d_1+k$ \textbf{do}
        \STATE \quad \quad $\mathbb{M}\leftarrow \mathbb{M}_0$
        
        \STATE \quad Update $\mathbb{M}_0$, $\mathbb{M}_1$ to fit $\mathbb{D}_0$, $\mathbb{D}_1$ respectively
        \STATE \quad Calculate the posterior $p(y|x, \mathbb{D})$ with $\mathbb{M}$
        \STATE \quad $x_i \leftarrow argmax_x(EHVI(\mathbb{A}, p(y|x, \mathbb{D})))$
        \STATE \quad Evaluate $x_i:y_i \leftarrow f(x_i)$
        \STATE \quad Update the prior: $\mathbb{D} \leftarrow \mathbb{D} \cup (x_i, y_i)$
        \STATE \quad Calculate the Pareto set: $\chi \leftarrow$ Pareto set of $\mathbb{D}$

        \STATE \textbf{end for}
        
        \STATE Return the current Pareto set $\chi$.
    \end{algorithmic}
\end{algorithm}

Algo.~\ref{alg:mfmobo} outlines the overall procedure of our proposed MFMOBO methodology. Initially, we sample and evaluate $d_0$ and $d_1$ points to build two prior datasets $\mathbb{D}_0$ and $\mathbb{D}_1$, corresponding to the evaluation functions of $f_0$~(high fidelity) and $f_1$~(low fidelity). We then iteratively explore the design space with $N_1$ trials of $f_1$ and $N_0$ trials of $f_0$ in total. In each iteration, we first update the surrogate models~($\mathbb{M}_0$, $\mathbb{M}_1$) to fit the datasets, and then calculate the posterior distribution of the entire space. In \name, we jointly optimize the performance and power of WSCs on LLM workloads to find the Pareto optimal designs. Thus we use the hypervolume as the optimization indicator and iteratively select design points with the maximum Expected Hypervolume Improvement (EHVI). Specifically, during the first $N_1$ trials, we evaluate the points with $f_1$ and predict the next promising point $x_i$ with the surrogate model $\mathbb{M}_1$. For the following k iterations from $N_1-d_1$ to $N_1-d_1+k$, we switch to $f_0$ as the evaluation function while still predicting the EHVI of candidate design points with $\mathbb{M}_1$. This process utilizes the information of the low-fidelity surrogate model to guide the initial search of high-fidelity optimization, and the selected design points are added to dataset $\mathbb{D}_0$ for surrogate model $\mathbb{M}_0$ to update. Finally, we switch to $\mathbb{M}_0$ for EHVI calculation during the rest of the iterations.

In \name, we utilize the Gaussian Process (GP) as the surrogate model. For hypervolume calculations, we define the reference point with a throughput of 0 and the power as the peak power threshold of the WSC system. During our evaluation process, due to the considerable time consumption of CA simulation, we incorporate only the analytical model and GNN-based evaluation as low-fidelity and high-fidelity evaluation functions, respectively, within the iterations of MFMOBO. By substituting the evaluation function within Algo.~\ref{alg:mfmobo}, our algorithm also supports further exploration based on CA simulation results.

\section{Experiment}

\subsection{Experimental Setup}

\noindent{\textbf{Design Space Setup.}} 
To determine the architecture parameters for our proposed WSC designs, we select candidate values as listed in Table~\ref{tab:candidate}. We consider the clock frequency of 1 GHz and set the peak power threshold of 15 kW~\cite{pal2019architecting} per wafer. We assume the maximum operating temperature of 105$^{\circ}$C, which is the junction temperature for WSC with two heat sinks~\cite{pal2019architecting}. For SRAM, we assume a voltage of 0.9V and use the ssg process for area and power estimation. In the NoC design, routers operate at 1V and support 8 input virtual channels (vc) and 4 buffers per vc, without physically shared buffers. We consider the area overhead for inter-reticle communication as $3900 \mu m ^ 2 / Gbps$ for RDL, and $1300 \mu m ^ 2 / Gbps$ for offset exposure. For stacking DRAM, TSV size and pitch are set to $5 \mu m$ and $15 \mu m$~\cite{samal2016monolithic}, with $1Gbps / TSV$ DRAM bandwidth. To model the trade-off between stacking memory capacity and bandwidth, we select several existing configurations and perform linear fitting. We set the area constraint of $26mm \times 33mm$ for reticles~\cite{reticle_limit} and consider $12-inch$ wafer with $215mm \times 215mm$ available area. All the area and power data are scaled to $14nm$ according to the scaling factors in~\cite{villa2014scaling}.

\begin{table}[]
\centering
\caption{Candidate Values for WSC Architecture Parameters}
\label{tab:candidate}
\resizebox{\columnwidth}{!}{%
\begin{tabular}{ll|ll}
\hline
\textbf{Core}                   & \textbf{}           & \textbf{Reticle}                        & \textbf{}                 \\ \hline
\multicolumn{1}{l|}{dataflow}     & WS, IS, OS          & \multicolumn{1}{l|}{inter\_reticle\_bw}  & 0.2-2 ($\times$ Bisection BW) \\
\multicolumn{1}{l|}{mac\_num}     & 8-4096              & \multicolumn{1}{l|}{stacking\_DRAM\_bw}  & 0.25-4 (TB/s/100$mm^2$)       \\
\multicolumn{1}{l|}{buffer\_size} & 32-2048 (KB)        & \multicolumn{1}{l|}{stacking\_DRAM size} & 8-40 (GB)                     \\ \cline{3-4} 
\multicolumn{1}{l|}{buffer\_bw} & 32-4096 (bit/cycle) & \textbf{Wafer}                          &                           \\ \cline{3-4} 
\multicolumn{1}{l|}{noc\_bw}      & 32-4096 (bit/cycle) & \multicolumn{1}{l|}{integration\_style}  & Die Stitching / InFO-SoW      \\
\multicolumn{1}{l|}{}           &                     & \multicolumn{1}{l|}{inter\_wafer\_bw}   & 100GB/s/Network Interface \\
\multicolumn{1}{l|}{}           &                     & \multicolumn{1}{l|}{off\_chip\_mem\_bw} & 160GB/s/MEM Controller    \\ \hline
\end{tabular}%
}

\end{table}

For yield modeling, we set a yield requirement of 0.9, and consider the average defect density as $D_{0} = 0.1/cm^2$~\cite{irds}. The yield loss rate and the maximum distance of influence are set to $0.1$ and $1mm$ for stress holes. Given the variability in yields across different core locations, we employ Monte Carlo sampling to estimate the yield of the reticle with redundant cores. For redundancy-based yield enhancement with minimum performance and area overhead, we refer to the design of Cerebras~\cite{lauterbach2021path} to add extra connections in each row of the core array. These additional connections can be dynamically configured to reroute data and computation when a core experiences a fault, enabling seamless and efficient core replacement. For the consideration of the KGD technique, we directly take the reticle yield as the yield for WSCs when employing inFo-SoW integration, while further calculating the yield for die-stitching WSCs with reticle yields.

\noindent{\textbf{LLM Benchmarks.}} We select a wide range of LLMs and scale the number of attention heads, hidden dimensions, and layers according to the setups in Megatron-LM~\cite{narayanan2021efficient}, GPT3~\cite{gpt3-1} and Zero-Infinity~\cite{rajbhandari2021zero}. We consider a fixed sequence length of 2048 and perform activation checkpoint with a granularity of 2 layers. For inference, we consider KV cache optimization and assume a constant sequence length of 2048 for both input and output, with a batch size of 32. During the DSE process, we set the total area of the WSCs to be consistent with that of the corresponding number of GPUs.

\noindent{\textbf{GNN Training Setup.}} To generate the dataset for GNN model training and validation, we randomly select a series of WSC configurations and LLM benchmarks. We generate the execution graph with Workload Compiler and collect the communication traces by evaluating the benchmark workloads with CA simulation. By meticulously tracking the transmission of packets within the NoC, we construct the transmission feature vectors, which serve as the regression targets for each sample in our dataset. Overall, our generated dataset comprises a total of 3000 samples. 

\subsection{Performance Model Verification}

We validate our performance model using publicly available TensorRT-LLM results for H100 and H200 GPUs \cite{tensorRT}, as well as a real-world LLM processing cluster. We configure our architectural design space to represent both the NVIDIA GPU clusters and the LLM processing cluster accurately. For GPU clusters, we disregard mesh-based NoC topology and rely on empirical utilization metrics for evaluating compute and communication performance. For the real-world LLM processing cluster, we profile the performance of compute operators and focus on calibrating communication latency. Additionally, we profile the theoretical bandwidth between devices using typical traces. Experimental results across various LLM workloads demonstrate an average error of less than 10\% between our evaluation results and the ground truth. The errors may come from inaccuracies in utilization estimates, as well as overheads related to kernel launches and the software stack. Despite these factors, our performance model remains sufficiently accurate to meet the needs of early-stage DSE. Furthermore, DSE focuses more on relative relationships rather than absolute values. Therefore, in subsequent validations, we further analyze the ordinal association of the evaluation results.

\begin{figure}[t]
  \centering
  \includegraphics[width=\columnwidth]{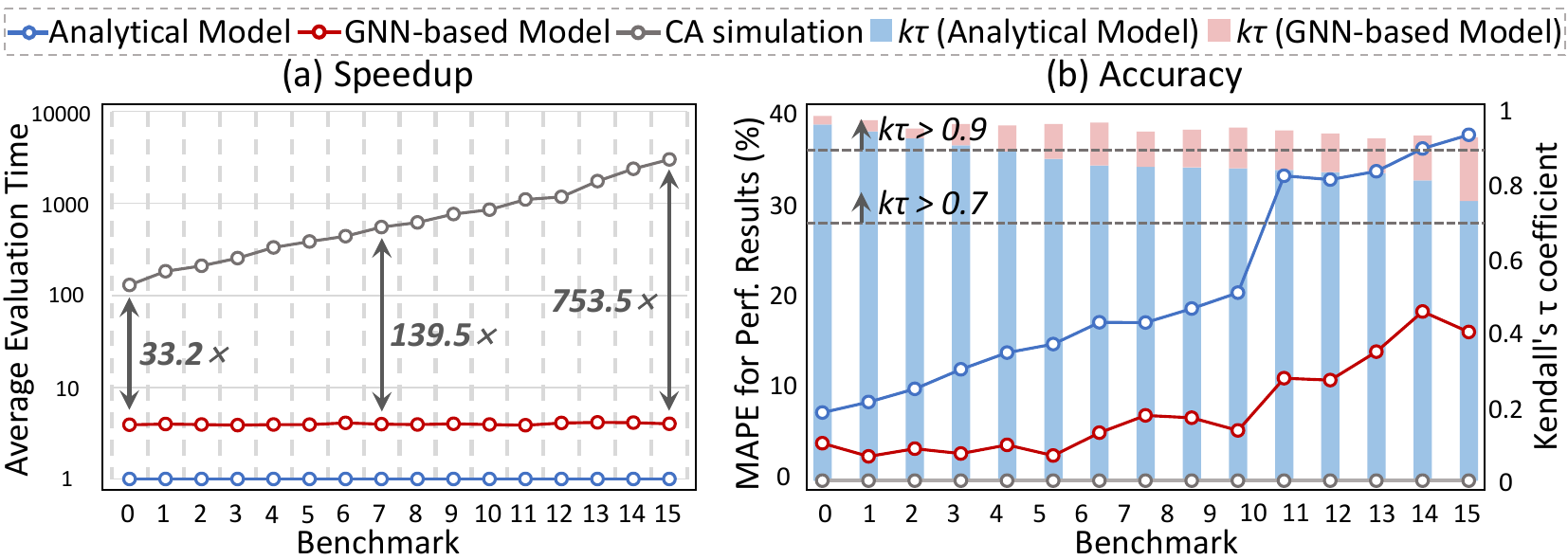}
  \caption{Evaluation Speedup and Accuracy Comparison}
  \label{fig:verify}
\end{figure}

Fig.~\ref{fig:verify} compares the speedup and accuracy of our evaluation models across 16 growing-scale benchmarks where CA simulation is refered to as the ground truth. As depicted in Fig.~\ref{fig:verify}~(a), the evaluation time with CA simulation increases significantly with the growing scale of workloads. In contrast, both the analytical model and GNN-based evaluation model maintain stable evaluation times across different workloads. Compared to CA simulation, GNN-based evaluation methodology demonstrates a speedup ranging from 33.2$\times$ to 753.5$\times$, with an average speedup of 220.2$\times$. Fig.~\ref{fig:verify}~(b) shows the evaluation accuracy, where CA simulation results are referred to as the ground truth. By aggregating congestion-relevant information, GNN-based evaluation outperforms the analytical model on all benchmarks, achieving an average error rate of 7.44\% compared to 20.29\% for the analytical model. To further validate the impact of errors from the two evaluation models on DSE results, we calculate and analyze Kendall's $\tau$~(KT) coefficient of the analytical model and GNN-based evaluation in comparison to CA simulation results. KT coefficient measures the ordinal association between the two evaluation methodologies and ground truth. In practice, a KT correlation above 0.9 is efficient for early-stage exploration of a design, while a KT correlation above 0.7 can generally benefit multi-fidelity optimization. As depicted in Fig.~\ref{fig:verify}~(b), the KT coefficient of the analytical model gradually decreases from 0.9 to 0.73 with increasing model size, while for GNN-based evaluation, the KT coefficient remains consistently above 0.9. Although both the analytical model and GNN-based evaluation model deviate further from the ground truth as the workload scale increases, the KT coefficient results indicate that they can still be efficient for the DSE process.




\subsection{Explorer Efficiency Analysis}

\begin{figure}[t]
  \centering
  \includegraphics[width=\columnwidth]{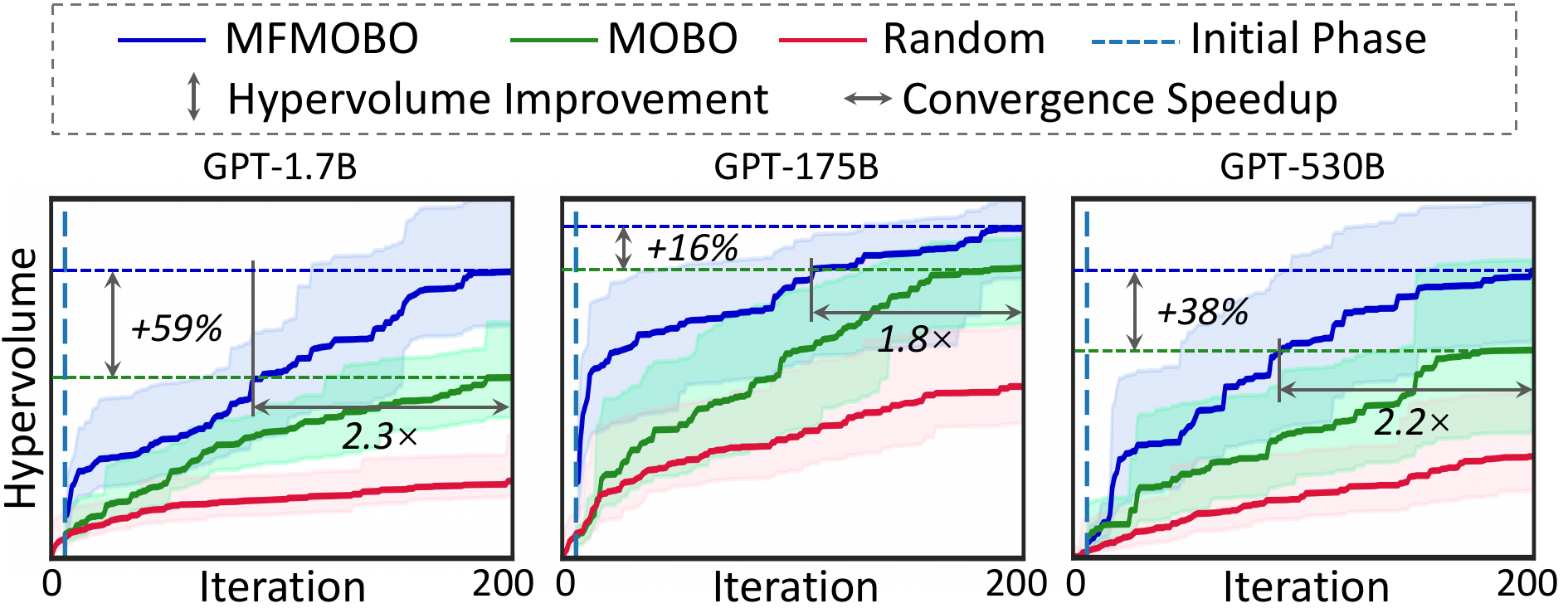}
  \caption{Optimization Results Comparison}
  \label{fig:convergence}
\end{figure}

To assess the efficiency of our proposed MFMOBO methodology, we compare MFMOBO with random search and traditional multi-objective Bayesian optimization~(MOBO). We employ the GNN-based evaluation model for both random search and MOBO while applying the analytical model together with GNN in MFMOBO. Both random search and MOBO iterate 200 times, whereas MFMOBO performs 100 iterations for initialization with the low-fidelity analytical model and takes the remaining time for high-fidelity iteration. For both MOBO and MFMOBO, we select an initial set with 6 design points. Fig.~\ref{fig:convergence} presents the optimization results for GPT-1.7B, GPT-175B, and GPT-530B, with similar trends observed for all benchmark LLMs. All experiments are repeated 10 times to calculate the average hypervolume improvement during the optimization iterations. Compared with vanilla MOBO, our proposed MFMOBO achieves an average 2.1$\times$ faster convergence to the same hypervolume on all LLM benchmarks, and on average 42\% hypervolume improvement within the same iteration time, resulting in optimal designs that closely approximate the real Pareto frontier. By leveraging the advantages from enhanced convergence and accelerated evaluation speed, \name\ attains an overall speedup of over 400$\times$ compared with MOBO with CA simulation.

\section{Case Studies and Insights}

\subsection{Core Granularity Tradeoffs}\label{sec:core_granularity}

\begin{figure}[t]
  \centering
  \includegraphics[width=\columnwidth]{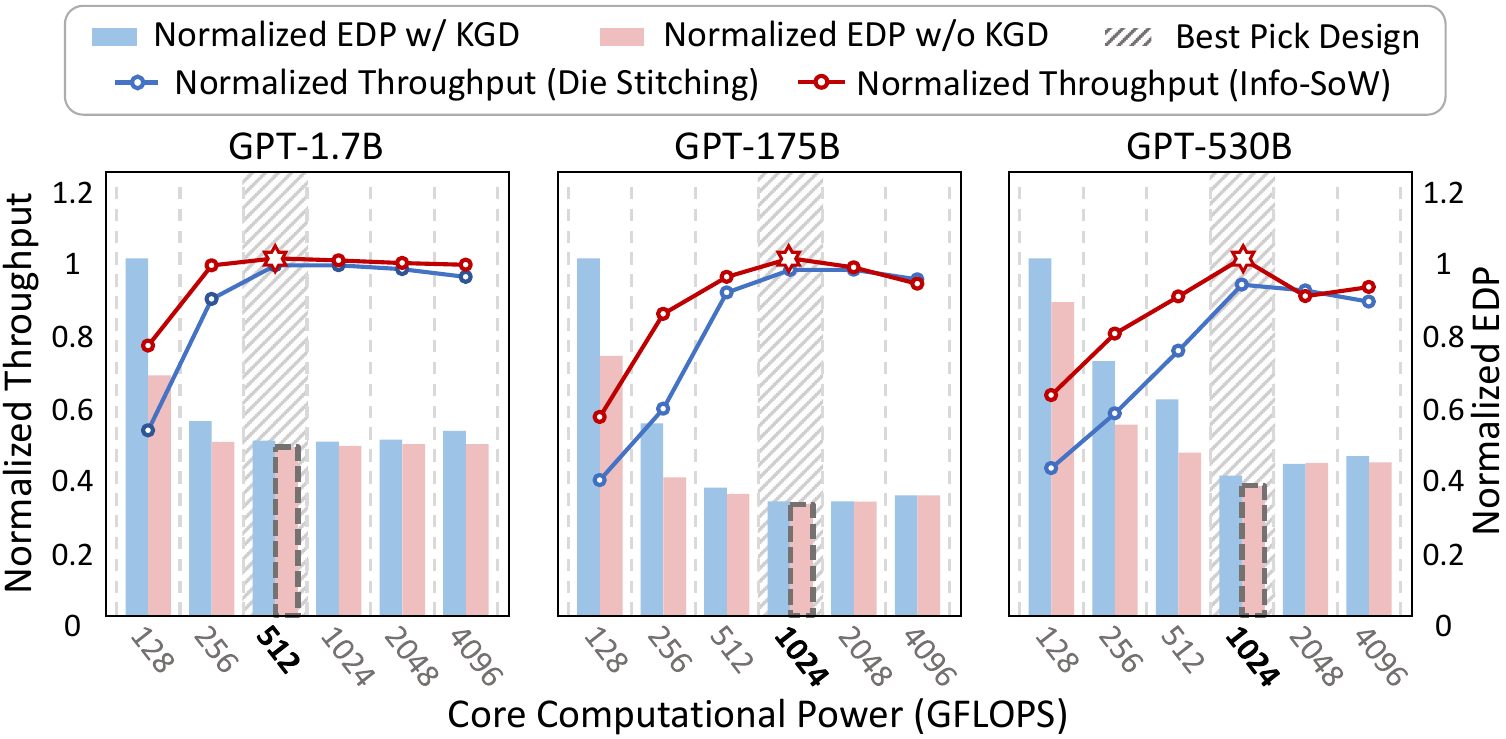}
  \caption{Core Granularity Exploration Results}
  \label{fig:core_granularity}
\end{figure}

In the design of WSCs for LLM workloads, the granularity of the core emerges as a critical factor. We define the computational power of the core (in FLOPS) as a representative measure of core granularity and explore combinations of all other parameters (i.e., SRAM capacity, NoC bandwidth, etc.) to find the optimal design in performance and energy consumption. Fig.~\ref{fig:core_granularity} illustrates the trend of training throughput and Energy-Delay Product (EDP) concerning the computational power of cores for LLMs of different scales. 


Generally, large cores tend to exhibit better area and energy efficiency than small cores, since they replace NoC-based interconnections between small cores with fixed datapaths. Besides, large cores can reduce the number of nodes in the NoC, simplifying routing complexity and alleviating NoC traffic. However, large cores face challenges related to utilization, module efficiency, and yield considerations.

\noindent\textbf{Utilization.} Small cores generally offer more scheduling flexibility, leading to better resource utilization. This is because within a core, the computation is operated in fixed dataflow, which relies on the parallelism of specific dimensions to fully utilize all processing units (MACs). However, in LLM workloads, the dimensions of the operators are generally large enough to effectively utilize computational resources within individual cores across various dataflows, as long as the SRAM capacity meets the requirement of data reuse. Within our candidate range of core granularity, the design of large cores does not pose utilization challenges. 

\noindent\textbf{Module Efficiency.} Effectively designing large cores with high computational power necessitates a corresponding increase in SRAM capacity and NoC bandwidth. However, both large-capacity SRAM designs and high-bandwidth NoC routers are inefficient in area and energy consumption, which may contribute to the decrease in throughput and EDP when core computational power surpasses a certain threshold. 

\noindent\textbf{Yield Consideration.} Larger cores exhibit lower yields and incur higher overhead due to redundant cores and extra connections. Consequently, the increase in core granularity may lead to a decline in both performance and energy efficiency.

In our experimental setup, the optimal computational power for cores falls within the range of 512G-1TFLOPS.

\noindent\fbox{%
  \parbox{0.47\textwidth}{%
\textbf{Takeaway 1}: 
For LLM workloads, WSC can be designed with large cores, while considering utilization, module efficiency and yield requirements.
}
}

\subsection{Integration Style Tradeoffs}

In Fig.~\ref{fig:core_granularity}, we also compare the performance under integration styles of die stitching and Info-SoW. Compared to die stitching integration, Info-SoW introduces a larger area and power overhead for inter-reticle communication. However, since know-good-die~(KGD) technology can only be applied to InFo-SoW,  InFO-SoW can always provide better performance and power consumption than die stitching due to this flexibility in yield guarantee. 
Meanwhile, for both integration styles, WSC designs should balance between the benefits of enhanced inter-reticle bandwidth and the introduced area overhead affecting computational resources. In our experimental setup, optimal inter-reticle bandwidths generally fall between 0.5$\times$ and 1$\times$ of the bisection bandwidth of a reticle.

\noindent\fbox{%
  \parbox{0.47\textwidth}{%
\textbf{Takeaway 2}: 
Info-SoW with KGD can outperform offset exposure due to less yield overhead, despite the larger area overhead for inter-reticle communication.
}
}

\subsection{Reticle Granularity Tradeoffs}



\begin{figure}[t]
  \centering
  \includegraphics[width=\columnwidth]{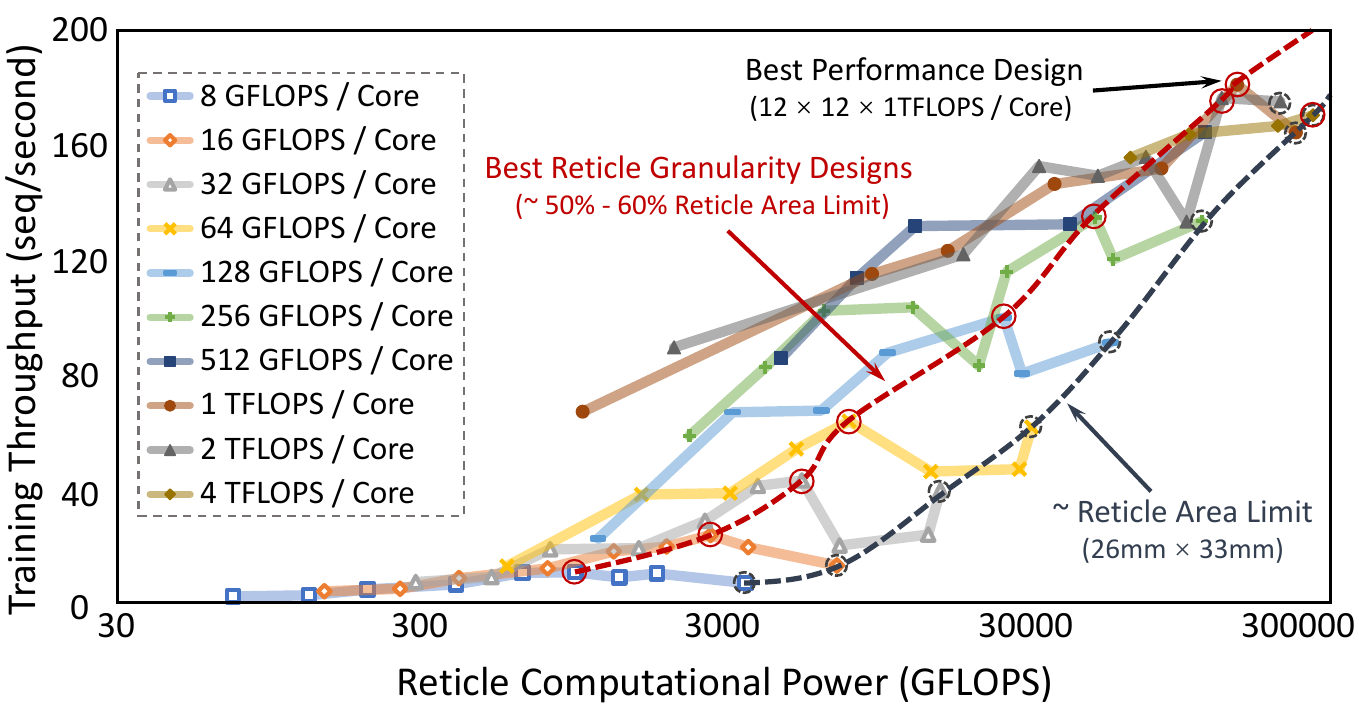}
  \caption{Reticle Granularity Exploration Results}
  \label{fig:reticle_granularity}
\end{figure}

Similar to core granularity, we refer to the peak computational power of a reticle as the measure of reticle granularity. In exploring the design of reticle granularity, we generate different configurations of cores, as well as feasible core array sizes under the reticle area constraint to calculate the computational power of the reticle. Fig.~\ref{fig:reticle_granularity} shows the optimal training throughput under given reticle granularities for GPT3 on WSC systems. As shown in Fig.~\ref{fig:reticle_granularity}, the computational power of a reticle varies from tens of GOPS to hundreds of TOPS. Overall, larger reticle granularity tends to have better performance, since inter-reticle communication introduces higher overhead in latency, area, and energy consumption compared to NoC. Meanwhile, larger reticle granularity can lead to reduced TP size, which reduces the amount of data transmission in collective communications. 

To further analyze the results, we cluster the designs with the same core granularity, where different reticle granularity reflects the core array size. We label the optimal reticle design under each core granularity, as well as the largest scale design within the reticle area constraint. It can be observed that the best performance designs are mostly not associated with the largest array size that approaches the area limit of reticles. As the number of cores increases, the overhead of redundant cores and extra connections rises to meet the yield requirement, which can impact the achievable computational power under the same area.


In our experiment setup, the performance optimal reticle granularity for GPT3 is 144TFLOPS, with a core array size of $12 \times 12$ and a core granularity of 1TFLOPS. Notably, the optimal reticle granularity design typically occupies 50\%-60\% of the reticle area limit across different core granularities.

\noindent\fbox{%
  \parbox{0.47\textwidth}{%
\textbf{Takeaway 3}: 
WSC reticle scale should trade between the redundancy ratio of cores and the overhead of inter-reticle connections, instead of going for the reticle limit.
}
}

\subsection{Speedup Analysis for LLM Inference}

\begin{figure}[t]
  \centering
  \includegraphics[width=\columnwidth]{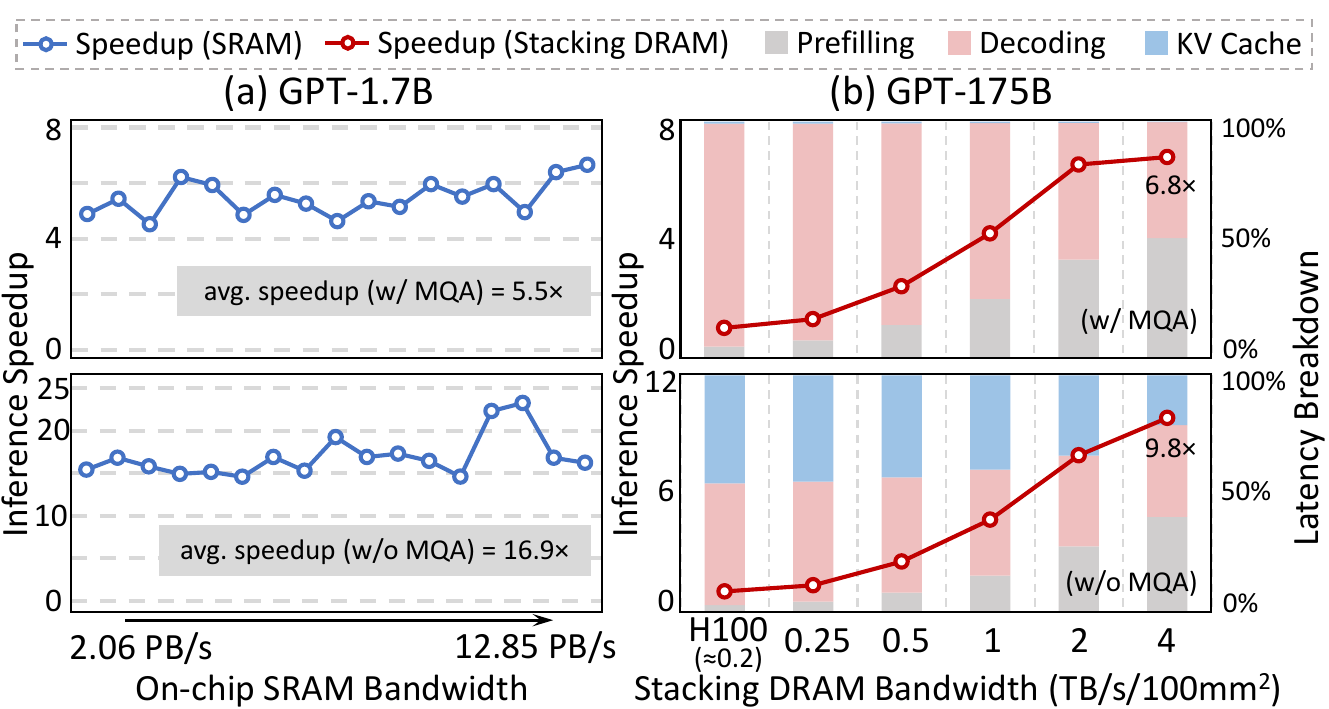}
  \caption{LLM Inference Speedup Comparison}
  \label{fig:inference_tradeoff}
\end{figure}

Fig.~\ref{fig:inference_tradeoff} depicts the inference speedup of WSCs over the H100 baseline~\cite{DGXH100} with the same area. 
Fig.~\ref{fig:inference_tradeoff}(a) shows the results on GPT-1.7B when all necessary data (weights, inputs, and KV-cache) are stored in the SRAM of WSCs. The x-axis illustrates the available on-chip SRAM bandwidth per WSC while meeting SRAM capacity requirements. WSC achieves an average speedup of 5.5$\times$ with multi-query attention~(MQA) optimization, and 16.9$\times$ without MQA. Notably, the LLM decoding stage is memory-bound on GPUs, particularly under small batch sizes, which leads to significant under-utilization of compute resources. When the SRAM capacity of the WSC is sufficient, a large SRAM bandwidth ensures optimal utilization of computational resources and benefits the load of KV-cache. However, increasing either capacity or bandwidth results in an increase in the SRAM area, which may influence the area of computational resources. This explains the variation of speedup and further motivates our exploration.



Fig.~\ref{fig:inference_tradeoff}(b) shows the inference speedup and latency breakdown of GPT-175B with stacking DRAM. Within the stress constraint, stacking DRAM bandwidths are varied from 0.25 to 4 TB/s/100$mm^2$. As for comparison, the HBM bandwidth of H100 is approximately 0.2 TB/s/100$mm^2$. The increased stacking DRAM bandwidth benefits both the decoding stage and KV-cache access, while achieving up to 6.8$\times$ speedup over GPU baseline with MQA and 9.8$\times$ without MQA. However, as bandwidth increases, the area of the TSV region grows, diminishing the effective computational power of the WSC and affecting the yield. This can lead to an increase in the latency of the prefilling stage, which may become the performance bottleneck. Meanwhile, the reduction in stacking memory capacity can result in increased inter-reticle communication, necessitating careful design of communication bandwidth to ensure the overall performance improvement. 

\noindent\fbox{%
  \parbox{0.47\textwidth}{%
\textbf{Takeaway 4}: 
The high bandwidth of both on-chip SRAM and stacking DRAM can efficiently speedup the inference for LLMs with different parameter scales, while the resulting capacity reduction can be alleviated through the efficient inter-reticle communication of WSCs.
}
}

\subsection{Heterogenous Improvement}

\begin{figure}[t]
  \centering
  \includegraphics[width=\columnwidth]{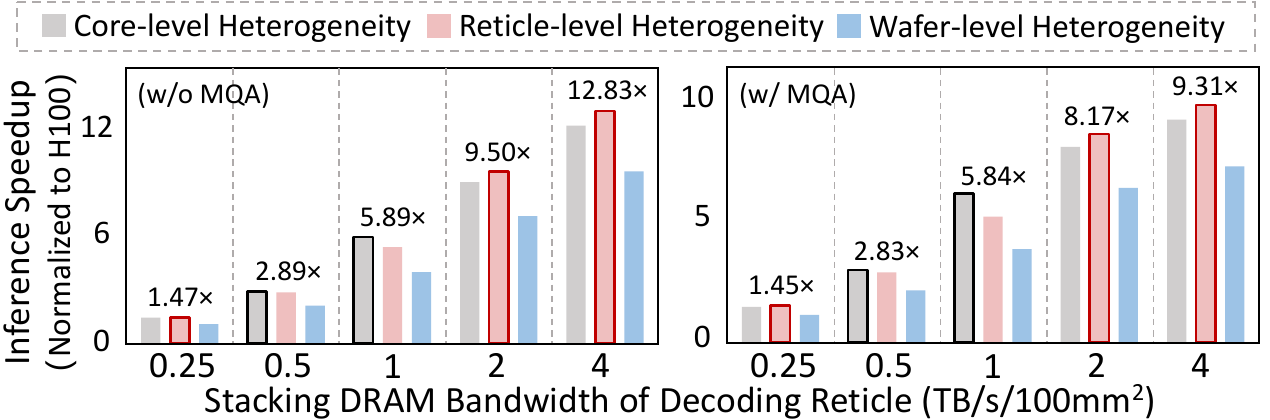}
  \caption{LLM Inference Speedup with Heterogenous. The searched best-performing configuration for the decode stage is 512GFLOPS with 32KB SRAM per core, 9$\times$9 cores per reticle, with 0.6TB/s inter-reticle bandwidth, and 10$\times$7 reticles per WSC.}
  \label{fig:inference_hetero}
\end{figure}

Fig.~\ref{fig:inference_hetero} depicts the speedup of GPT-175B inference with various levels of heterogeneity. To address hardware heterogeneity, we separately optimize the performance of the prefill and decode stages during the exploration process, while considering the inter-stage KV-cache transfer overhead. By adjusting the resource allocation between the two stages, we can achieve the optimal overall throughput. As shown in Fig.~\ref{fig:inference_hetero}, with the same stacking memory bandwidth for the decode stage, heterogeneous design can yield higher inference speedup. This is primarily attributed to the augmentation of hardware computational power and resource utilization during the prefill stage. Due to the impact of increased stacking memory bandwidth on the effective area and yield of reticles, the computational power of reticles optimized for the prefill stage can exceed that of reticles optimized for the decode stage by over $1.6\times$ under the same area. We also compare the performance improvements resulting from different levels of heterogeneity in Fig.~\ref{fig:inference_hetero}, and mark the optimal heterogeneous granularity for each configuration. 
Due to limited inter-wafer bandwidth, KV-cache transfer can become the bottleneck when applying wafer-level heterogeneity.
For core-level heterogeneity, the flexible scheduling and resource allocation of compute-intensive and memory-intensive operators on the same reticle enables higher resource utilization. However, it also introduces an increased volume of both intra-reticle and inter-reticle transmissions, as well as overhead in compilation and control. Reticle-level heterogeneity strikes a favorable balance between these factors, although it may pose challenges to existing integration technologies.

\noindent\fbox{%
  \parbox{0.47\textwidth}{%
\textbf{Takeaway 5}: 
Reticle-level heterogeneity can provide the best tradeoff for LLM inference between hardware utilization of both stages and introduced overhead.
}
}

\subsection{DRAM Approach Analysis}

In Fig.~\ref{fig:dse_results}, we visualize the overall design space for GPT-175B training. Each point represents a sampled design configuration during our iterative exploration process, where red points correspond to designs with traditional off-chip DRAM approach and blue points correspond to those utilizing stacking DRAM. From the comparison of the Pareto frontiers, it can be observed that WSCs can benefit from stacking DRAM in both performance and power efficiency. Traditional off-chip DRAM necessitates access through the memory interfaces around the wafer, raising the transfer pressure of inter-reticle communications. Long-range DRAM-access-induced data transfer from the WSC edge can become the performance bottleneck, introducing additional power consumption, with implications on the available inter-reticle bandwidth. 

\noindent\fbox{%
  \parbox{0.47\textwidth}{%
\textbf{Takeaway 6}: 
Stacking memory allows for better scaling than off-chip memory due to its higher bandwidth, improved power efficiency, and reduced communication overhead.
}
}


\subsection{Design Space Exploration Results}\label{sec:dse_results}

\begin{figure}[t]
  \centering
  \includegraphics[width=\columnwidth]{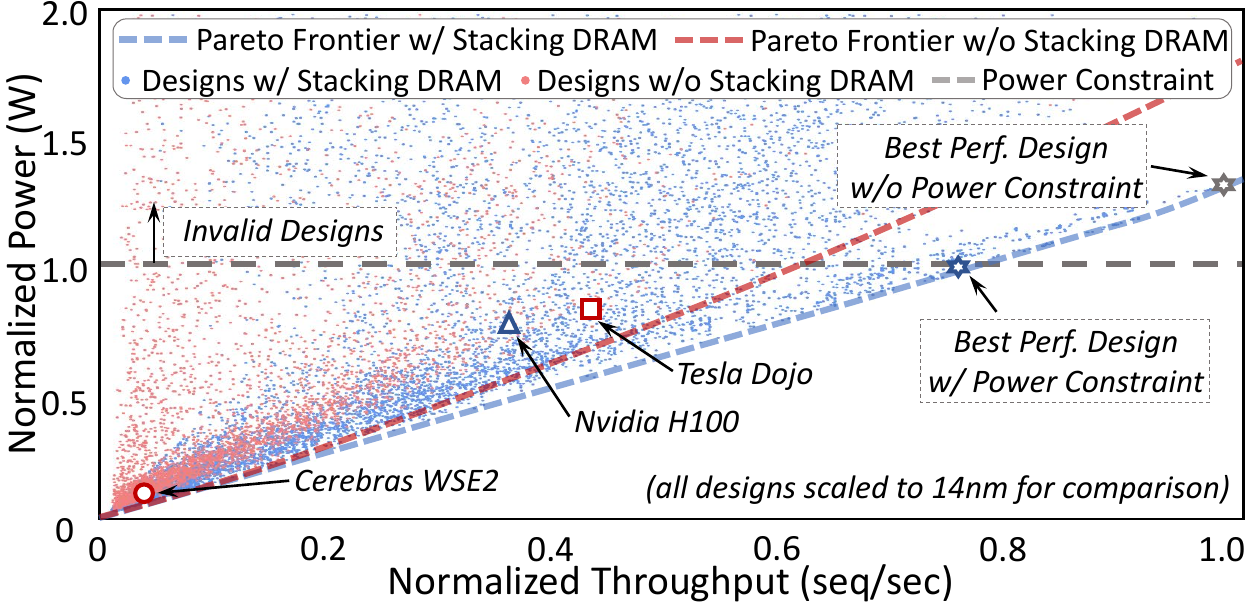}
  \caption{Design Space for GPT-175B Training. The searched best-performing configuration is 1TFLOPS with 128KB SRAM per core, 12$\times$12 cores per reticle, with 1.5TB/s inter-reticle bandwidth, and 9$\times$6 reticles per WSC.}
  \label{fig:dse_results}
\end{figure}

We search for Pareto Optimal WSC configurations for both training and inference across various LLM benchmarks: Fig.~\ref{fig:inference_hetero} and Fig.~\ref{fig:dse_results} present two typical configurations for GPT-175B inference and training. We calculate the area of these configurations with our area model, and find that most of these configurations occupy around 50\% - 70\% of the $215mm \times 215mm$ wafer limit. Specifically, the heterogeneous reticles in Fig.~\ref{fig:inference_hetero} account for 55.2\% (with 16.3\% for decode stage reticles), while the design in Fig.~\ref{fig:dse_results} occupies 69.2\%.

\noindent\fbox{%
  \parbox{0.47\textwidth}{%
\textbf{Takeaway 7}: 
The scaling up of a single WSC does not always need to reach the wafer limit for LLM workloads.}
}

To demonstrate the effectiveness of \name\ framework, we compare the performance and power of searched Pareto optimal WSCs with existing designs including H100~\cite{DGXH100}, Cerebras WSE2~\cite{lie2022cerebras} and Tesla Dojo~\cite{chang2022dojo}. All comparisons are made under the same area, with both area and power values for existing designs scaled to 14nm. For H100, we ignore yield requirements and the area overhead for NVLink Serdes. For the training tasks across our LLM benchmarks, \name\ can find Pareto optimal WSC designs that outperform the H100 cluster by an average of 62.8\% in performance (with the same or lower power) and 38.6\% in power (with the same or higher performance), thanks to the high bandwidth and low power inter-reticle communication, efficient stacking DRAM and effective compute utilization. Meanwhile, the results obtained by \name\ outperform Cerebras WSE2/Tesla Dojo by up to 73.7\%/46.5\% in performance and 42.4\%/31.7\% in power, respectively. In comparison with existing WSC designs, our searched Pareto optimal designs benefit from the proper selection of core granularity, inter-reticle bandwidth, and the effectiveness of stacking DRAM. For LLM inference tasks, WSC designs improve up to 23.2$\times$/15.7$\times$ for performance and power with sufficient SRAM capacity, and up to 12.9$\times$/11.2$\times$ with stacking DRAM.


\section{Conclusion}

In this paper, we propose \name, a DSE framework designed to facilitate high-efficiency WSC design for LLM workloads. We construct a comprehensive design space for WSCs considering various design constraints. We propose hierarchical evaluation methodologies for efficient evaluation of design points and design a MFMOBO methodology to efficiently explore the WSC design space. 
Experimental results demonstrate that \name\ significantly enhances the efficiency of the DSE process, and the searched Pareto optimal WSC configurations can outperform existing designs. Furthermore, we analyze the exploration results and provide insights to design efficient WSCs for LLMs.


\bibliographystyle{IEEEtranS}
\bibliography{refs}

\end{document}